\documentclass[letterpaper,3p,times,twocolumn]{elsarticle}

\usepackage{ecrc}

\volume{00}

\firstpage{1}

\journalname{Nuclear Physics B Proceedings Supplement}

\runauth{A.~Gro{\ss} et al.}

\jid{nuphbp}

\jnltitlelogo{Nuclear Physics B Proceedings Supplement}

\usepackage{amssymb}
\usepackage[figuresright]{rotating}

\begin{document}

\begin{frontmatter}



\dochead{}

\title{Atmospheric Neutrino Oscillations in IceCube}

\author[munich]{A.~Gro{\ss}}
\author[IC]{ on behalf of the IceCube collaboration}

\address[munich]{Technische Universit\"at M\"unchen, D-85748 Garching, Germany}
\address[IC]{for full author list see http://icecube.wisc.edu/collaboration/authors/current}

\begin{abstract}
We present the results of an analysis of data collected by IceCube/DeepCore in 2010-2011 resulting in 
the first significant detection of neutrino oscillations in a high-energy neutrino telescope. 
A low-energy muon neutrino sample (20-100 GeV) containing the oscillation signal
was extracted from data collected by DeepCore. 
A high-energy muon neutrino sample (100 GeV -10 TeV) was extracted from IceCube data in order to constrain the systematic uncertainties. 
The non-oscillation hypothesis was rejected with more than $5\sigma$. We fitted the oscillation parameters $\Delta m^2_{23}$ and $\sin^22 \theta_{23}$ 
to these data samples. In a 2-flavor formalism we find 
$\Delta m^2_{23}= (2.5\pm0.6)\cdot10^{-3}$ eV$^2$ and $\sin^22 \theta_{23}>0.92$ while maximum mixing is favored. 
These results are in good agreement with the world average values.

\end{abstract}


\end{frontmatter}

% \linenumbers

\section{Introduction}
\label{intro}
Neutrino oscillation experiments have established that neutrino flavor and mass eigen states do mix\cite{Beringer:1900zz}. So far, solar and long-baseline
reactor neutrino experiments have measured the mass-mixing parameters ($\delta m^2, \theta_{12}$) in the $\nu_e \rightarrow\nu_e$ channel 
(electron neutrino disappearance), while atmospheric and long-baseline accelerator experiments have measured ($\Delta m^2, \theta_{23}$) in the 
$\nu_\mu \rightarrow \nu_\mu$ channel (muon neutrino disappearance)\footnote{We here adopt a convention where $\delta m^2=m_2^2-m_1^2$ and\\ $\Delta m^2=m_3^2-(m_2^2+m_1^2)/2$.  }.

We use data collected from May 2010 to May 2011 by the IceCube neutrino telescope with its low-energy sub-detector DeepCore \cite{Collaboration:2011ym} to measure the atmospheric neutrino oscillation parameters.  
The IceCube Neutrino Observatory is a cubic-kilometer neutrino detector installed in the ice at
the geographic South Pole \cite{Achterberg:2006md}. It is based on the optical detection of secondary particles produced by neutrinos interacting in the ice or the bedrock below. 
Such charged particles emit Cherenkov light, which is detected by IceCube's optical sensors. 
These sensors are attached to 86 strings, which hold 60 sensors each. This corresponds to a standard vertical spacing of 17 m between sensors and a 
horizontal distance of 125 m between the strings. The DeepCore sub-detector consists of eight additional strings deployed in the center of IceCube and the 
surrounding IceCube strings. On the dedicated DeepCore strings, the sensors are concentrated in the cleanest deep ice, resulting in a denser $7$ m vertical 
spacing of sensors there. During the data taking period of this analysis, 79 detector strings were operational (IceCube-79) including six dedicated DeepCore strings.

We used a 2-flavor formalism to describe neutrino oscillations, neglecting 3-flavor effects and matter effects. In this formalism, 
the muon neutrino survival probability is given by 
\begin{equation}
P(\nu_\mu\rightarrow\nu_\mu)= 1 -\sin^2 2\theta_{23} \sin^2(1.27 \Delta m^2L/E) 
\end{equation}
with $L$ as the length of propagation of the neutrino in km and $E$ as the neutrino energy in GeV.

\section{Data sample}
\label{data}
\begin{figure}[htb!]
\begin{center}
\includegraphics[width=0.35\textwidth]{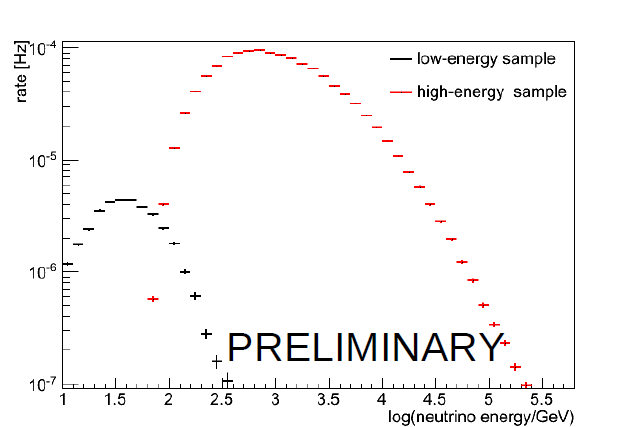}
\end{center}
\caption{\label{fig:energy}Distribution of the neutrino energy of atmospheric neutrinos in the low-energy 
(DeepCore) and in the high-energy (IceCube) sample.}
\end{figure}

We extracted two samples of upwards going neutrino events from data collected by IceCube-79, 
one at relatively high energies using data from the entire IceCube detector and one at lower energies selected in the 
DeepCore volume, 
rejecting backgrounds %( throughgoing neutrino-induced high-energy muons and downwards going muons from air showers) 
by an active veto \cite{SchulzEulerICRC}. 
Neutrino oscillations are expected to affect only the low-energy sample. 
The high-energy sample provided large statistics outside the signal 
region and served to constrain systematic uncertainties.

The directions of the muon tracks in the high-energy sample were reconstructed with the  standard  maximum 
likelihood muon track in IceCube \cite{Ahrens:2003fg}.
For low-energy events, the same method was applied as an initial step. 
As the hypothesis of a throughgoing track is not correct for these, the 
finiteness of the tracks was considered in a second step. 
The track length and the start/stop points of the track were determined as well as 
the likelihood whether the track is starting and/or stopping in the detector \cite{SchulzEulerICRC}. 
Quality cuts like the number of unscattered photons and the track likelihood allowed for the rejection of misreconstructed downwards 
going muon background. 

\begin{figure}[htb!]
\begin{center}
\includegraphics[width=0.35\textwidth]{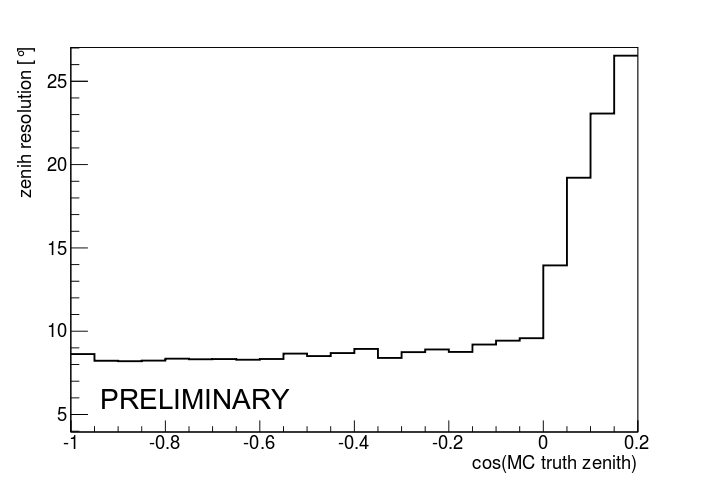}
\end{center}
\caption{\label{fig:zenithresolution}Zenith resolution (mean absolute difference between true and reconstructed track) for the low-energy sample as a function of true zenith.}
\end{figure}

In Fig.~\ref{fig:energy}, the neutrino energy distribution of the low-energy and the high-energy sample are shown.
The resolution of the reconstructed zenith angle is essential because the propagation length is proportional to cos(zenith). 
A variation of the zenith thus represents a variation of L/E. As displayed in Fig.~\ref{fig:zenithresolution}, a resolution of $8^\circ$ is achieved for the low-energy sample, 
independent from the zenith. 

A modified version of the atmospheric neutrino flux model derived by Honda et al.~\cite{Honda:2006qj} was used 
because recent measurements of the spectrum of charged cosmic rays in the energy range of 200 GeV to 100 TeV 
indicate a cosmic ray spectrum flatter than that assumed by Honda et al., see e.g. \cite{Yoon:2011aa}. 
We adjusted the resulting neutrino spectrum by hardening its spectral index by 0.05 to reflect these new measurements.

\section{Systematic uncertainties}
\label{systematics}
A covariance matrix in a $\chi^2$ fit was used to consider systematic uncertainties in the data analysis. In order to obtain
the most likely value of the individual sources of systematic errors, the pulls as defined in \cite{Fogli:2002pt} were used.
The following sources of systematic uncertainties were considered explicitly and propagated by Monte Carlo (MC) 
simulation to the final selection level:
\begin{itemize}
\item the absolute sensitivity of the IceCube sensors ($\pm10\%)$ and the relative efficiency of the more efficient DeepCore sensors ($1.35\pm0.03$)
\item the optical parameters (scattering, absorption) of the ice as detector medium: the uncertainty is estimated by 
the difference of the optical parameters obtained by the extraction methods \cite{Millenium} and \cite{Spice}
\item the absolute normalization of the cosmic ray flux ($\pm25\%$) and its spectral index ($\pm0.05$)
\item the uncertainty of the neutrino production rate in the atmosphere: the difference of calculations
by  \cite{Honda:2006qj} and \cite{Barr:2003gy} were used for $\nu_\mu$ and for $\nu_e$.
\end{itemize}
%The difference to the standard MC provides
%the correlated errors defining the covariance matrix.
\section{Results}

\label{results}
\begin{figure}[htb!]
\begin{center}
\includegraphics[width=0.35\textwidth]{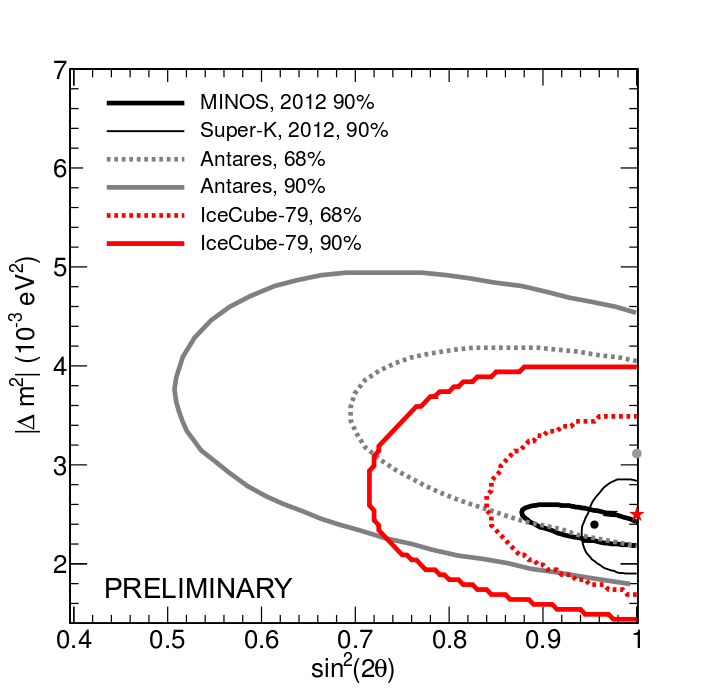}
\end{center}
\caption{\label{fig:contourlines} $68\%$ and $90\%$ CL contour line of the result of the IceCube-79 oscillation analysis in comparison with the 
results of Antares\cite{AdrianMartinez:2012ph}, Minos and SuperKamiokande\cite{SuperK_MINOS}.}
\end{figure}

%\begin{figure}[hb!]
%\includegraphics[width=0.3\textwidth]{Pulls_IC79_oscillation}
%\caption{\label{fig:pulls}Pulls on the systematic uncertainties at best fit value of $\Delta m^2=2.39e-3 eV^2$ and $sin^2(2 \theta_{23})=1$}
%\end{figure}

\begin{figure}[hb!]
\begin{center}
\includegraphics[width=0.35\textwidth]{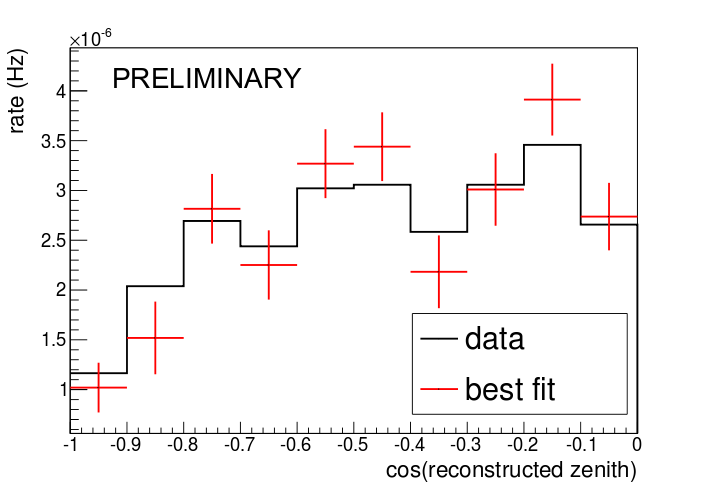}
\includegraphics[width=0.35\textwidth]{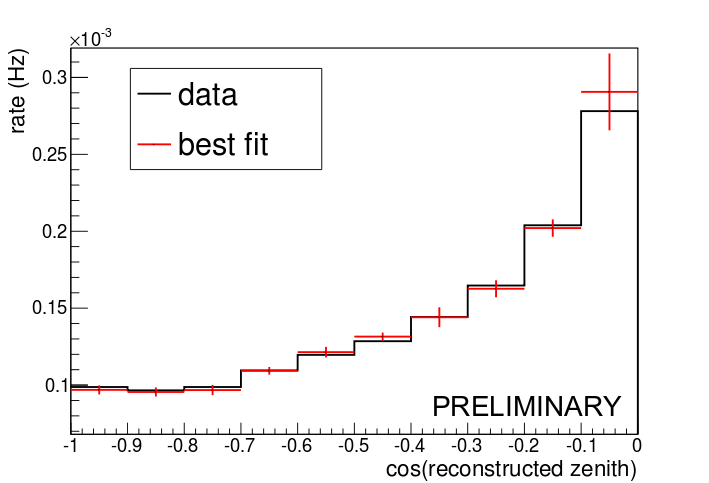}
\end{center}
\caption{\label{fig:bestfit}Data and MC expectation at best fit (physics parameters and pulls) in the low energy sample and for the high energy sample.} 
\end{figure}

During May 2010 to May 2011, we collected $318.9$ days of high quality data, excluding periods of calibration runs, partial detector configurations
and detector downtime. The low energy sample contained 719 events, while the high energy sample contained $39,638$ events after final cuts.
In a first step, we evaluated the $\chi^2$ for the data collected by IceCube for two different physics hypotheses: the standard oscillation scenario represented
by the world average best fit parameters and the non-oscillation case. With $\Delta\chi^2=30$, we rejected the non-oscillation hypothesis with a 
p-value of $10^{-8}$, corresponding to $5.6\sigma$. The significance was evaluated by a toy MC considering deviations 
from a $\chi^2$ distribution (both hypotheses do not correspond to the minimum $\chi^2$).

In a second step, the $\chi^2$ was evaluated as a function of the oscillation parameters. The best fit is given by $\Delta m^2_{23}=2.5\cdot10^{-3}$eV$²$ 
and $\sin^22 \theta_{23}=1$, with an absolute $\chi^2=11.3$ and 18 degrees of freedom (20 bins, 2 fitted parameters). 
The value of the absolute  $\chi^2$ corresponds to a goodness-of-fit p-value of $0.88$. This indicates
a good agreement of data to MC within the assumed uncertainties. All pulls on the systematic uncertainties are within the  $1\sigma$ uncertainty range.
The data as a function of zenith together with the statistical $1\sigma$ range of the MC expectation corrected for the 
pulls is shown in Fig.~\ref{fig:bestfit}. The result is in good agreement with other experiments, which measured the atmospheric oscillations with a high resolution 
at lower energies\cite{SuperK_MINOS}.

The two dimensional confidence regions of the oscillation parameters in this measurement were determined from the $\Delta\chi^2$ around the best fit with
two degrees of freedom ($68\%$ CL: $\Delta\chi^2=2.30$ and $90\%$ CL: $\Delta\chi^2=4.61$), see Fig.~\ref{fig:contourlines}.
The confidence regions of the individual parameters were determined by marginalization analogous to a profile likelihood method. We obtain 
$68\%$ CL intervals of $\Delta m^2_{23}= (2.5\pm0.6)\cdot10^{-3}$ eV$^2$ and $\sin^22 \theta_{23}>0.92$ using  $\Delta\chi^2$ with one degree of freedom.

The analysis of IceCube data presented here provided the first significant detection of atmospheric neutrino oscillations with a high-energy neutrino telescope. 
In future, a significant improvement of the resolution of IceCube on the atmospheric neutrino oscillations is expected by the inclusion of the reconstructed neutrino energy in the analysis, by the use
of new reconstruction methods which are more efficient at lower energies and by the inclusion of two additional dedicated DeepCore strings which started data 
taking in May 2011.



%% References

\nocite{*}
\bibliographystyle{elsarticle-num}
\bibliography{IC79_oscillations}






% \end{thebibliography}

\end{document}